# Helicity Maximization in a Planar Array of Achiral High-Density Dielectric Nanoparticles

Mina Hanifeh[1] and Filippo Capolino[2]

*Department of Electrical Engineering and Computer Science, University of California, Irvine, California 92697, USA*

**ABSTRACT**

We investigate how a periodic array composed of achiral isotropic high-refractive index dielectric nanospheres generates nearfield over the array surface reaching helicity density very close to its upper bound. The required condition for an array of nanospheres to generate "optimally chiral" nearfield, which represents the upper bound of helicity density, is derived in terms of array effective electric and magnetic polarizabilities that almost satisfy the effective Kerker condition for arrays. The discussed concepts find applications in improving chirality detection based on circular dichroism (CD) at surface level instead of in the bulk. Importantly the array would not contribute to the generated CD signal when used as a substrate for detecting chirality of a thin layer of chiral molecules. This eliminates the need to separate the CD signal generated by the array from that of the chiral sample.

## I. INTRODUCTION

Chirality is interpreted as lack of superimposable mirror images in an object [1,2]. Chirality is an outspread geometrical property: from helical orbitals of planets around the Sun to snail shells. It is also observed in building blocks of life, i.e., molecules [3,4]. Although many of natural molecules are found with one handedness, artificial substances which have become an indispensable part of our lives (e.g. medicines) mostly contain both enantiomers[1]. Conflicting effects of enantiomers on our bodies necessitate the separation of enantiomers and investigation of their properties[1,3].

There has been extensive research on determining handedness and strength of chirality in chiral materials [5–17]. One method is circular dichroism (CD) where transmitted powers of circularly polarized (CP) plane waves with opposite handedness through a layer of chiral medium with a thickness $d$ are used to reveal the material's chirality [12]. In 2010, it was suggested that replacing plane waves by structured lights with high helicity density improves chirality detection [5,6]. Subsequently, nearfields of various nanoantennas (NA) were proposed to enhance helicity in their nearfields [13,14,18–21]. In recent CD experiments, a chiral layer is deposited on an array of chiral plasmonic structures to enhance detection (see e.g., Refs. [7,22]). However, complexity of nearfields in vicinity of plasmonic structures [23–25] necessitates employing complex designs, such as an array of degenerate structures composed of dielectric cavities loaded by gold disks [26] to control helicity density in their vicinity. Recently, a NA composed of high refractive index material, such as a Si nanosphere, is proved to generate a nearfield with azimuthally symmetric helicity density of the same sign of that of the incident field [14]. Moreover, it has been shown that nearfield of an array of Si disks illuminated by a plane wave with circular polarization generates enhanced helicity density [27,28].

Here, we derive the required condition by an array of NAs to generate optimally chiral nearfield which represents the upper bound of obtainable helicity density with a given field energy density [16]. Indeed, we present a quantitative measure to investigate how close helicity density is to its upper bound. We prove that to have optimally chiral nearfield in an array of nanoparticles, all array elements are required to generate optimally chiral nearfield. This condition reduces to satisfying the Kerker condition by the effective electric and magnetic polarizabilities, which we call "effective" Kerker condition, when an array of achiral nanoparticles is considered. We rigorously analyze helicity density enhancement and distribution of nearfield of an array of spherical high-density dielectric NAs (made for example of Silicon or TiO$_2$). Importantly, each NA is isotropic and achiral [29] and the array does not contribute to CD signal when utilized as a substrate for chiral samples.

---

[1] M.hanifeh@uci.edu
[2] F.capolino@uci.edu

The organization of the paper is as follows. In Section II, we review the concept of helicity of electromagnetic fields as the chief quantity when dealing with chirality detection. In Sections III and IV, optical properties and helicity density of nearfields are, respectively, investigated in proximity of both an individual and an array of high-density NAs. Section V concludes our analysis.

## II. Helicity of fields

Helicity density for time harmonic electromagnetic fields reads [30–32] (when using the International System of Units)

$$h = \frac{1}{4c_0}(\epsilon_0 \mathbf{F}^* \cdot \nabla \times \mathbf{F} + \mu_0^{-1} \mathbf{A}^* \cdot \nabla \times \mathbf{A}) \quad (1)$$

where $\mathbf{A}$ and $\mathbf{F}$ are, respectively, the magnetic and electric vector potentials related to the electric $\mathbf{E}$ and magnetic $\mathbf{H}$ fields via $\mathbf{E} = -\nabla \times \mathbf{F}$ and $\mathbf{H} = \mu_0^{-1} \nabla \times \mathbf{A}$. Moreover, $c_0, \epsilon_0$ and $\mu_0$ are, respectively, the speed of light, permittivity, and permeability in vacuum, and the superscript "*" denotes complex conjugation. Note that fields with time dependence $\exp(-i\omega t)$, where $\omega$ is the angular frequency, are considered throughout the paper. Helicity is also closely related to optical chirality, a quantitative measure of chirality of electromagnetic fields, introduced by Tang and Cohen to represent the strength of field in detection of chirality of matter [5,6]. Indeed, for time-harmonic electromagnetic fields, the time-averaged optical chirality and helicity are proportional. Here, we adopt helicity as the measure to quantify fields' chirality, since it is shown that it is physically meaningful, rather than optical chirality [32].

After introduction of the chirality concept in electromagnetic fields and discussing its significance on chirality detection of matter, intensive research studies have been performed to propose structured lights, such as nearfields of NAs and optical beams, to enhance helicity density in comparison to plane waves with circular polarization [14,19,22,25,27,33–35]. Helicity enhancement depends on several factors which become clear when we employ the Coulomb gauge $\nabla \cdot \mathbf{A} = 0$ and $\nabla \cdot \mathbf{F} = 0$, and simplify Eq. (1) to

$$h = \frac{\Im\{\mathbf{E} \cdot \mathbf{H}^*\}}{2\omega c_0}. \quad (2)$$

Equation (2) shows that helicity is enhanced when: the magnitudes of electric and magnetic field components increase, their phase difference approaches $\pi/2$, and the electric field vector and the conjugate of the magnetic field vector become collinear. However, it has recently been proved that an upper-bound of helicity density of an electromagnetic field exists and is proportional to its energy density [16]. Fields which possess the maximum attainable helicity density are called *optimally chiral* and satisfy conditions

$$\mathbf{E} = \pm i\eta_0 \mathbf{H} \quad (3)$$

where $\eta_0$ denotes the intrinsic wave impedance of vacuum, and their helicity density takes the value of

$$h = \pm \frac{u}{\omega}, \quad (4)$$

where $u = (\epsilon_0|\mathbf{E}|^2 + \mu_0|\mathbf{H}|^2)/4$ is the time average energy density of the electromagnetic field. We note that the imaginary unit $i$ in Eq. (3) represents a phase difference of $\pi/2$ between the phasors of electric and magnetic field vectors and it means that the time varying $\mathbf{H}$ field is delayed or ahead the $\mathbf{E}$ field by a quarter of the time period $2\pi/\omega$. As discussed in [16], the helicity value in Eq. (4) constitutes the maximum (positive or negative) helicity any field with a given time average energy density $u$ may have, hence it represents a universal bound. When the field satisfies the "optimally chiral" relation it is possible to remove the local field information (that may not be available on most of the cases) form the expression of the dissymmetry factor $g$, to reveal chirality of matter. In other words, under this optimal condition the dissymmetry factor $g$, depends only on the properties of the chiral nano sample [8,16,36]. A simple example of field satisfying such "optimally chiral" relation is the circularly polarized plane wave. Note that although the significance of optimally chiral fields such as circularly polarized plane waves in CD experiments, is discussed in some publications [37,38], it is generally ignored when structured lights is employed to enhance the CD signal [5,6,18,23,28,39–45]. Another example is the azimuthally-radially polarized beam (ARPB) used for example in Refs. [10,36,46,47] that is a composition of an azimuthally polarized beam (APB) [48–52] and a radially polarized beam (RPB) [47,53,54]. Another remarkable example of optimally chiral fields, highly related to this paper, is the nearfield of a high-density dielectric (e.g. of Si or TiO$_2$) spherical particle with appropriate size and under suitable illuminations such as a Gaussian beam with circular polarization [8,36], and as an azimuthally-radially polarized beam [36,55] used, respectively, which provides also helicity nearfield enhancement with respect to that of an illuminating plane wave. In the following sections we discuss exactly that, i.e., helicity density enhancement in nearfield of a high-density NA and in an array composed of NAs.

## III. Helicity Analysis of Nearfield of a high-density dielectric NA

As a determinant factor in the electromagnetic response of an array, we devote this section to investigate the interaction of a spherical high-density NA with electromagnetic fields. Let us assume that a spherical NA with radius $a$ is located at the origin of the coordinate system and is irradiated by an incident field with electric and magnetic components $\mathbf{E}_{\text{inc}}$ and $\mathbf{H}_{\text{inc}}$, respectively, propagating along the $+z$ direction in free space (Figure 1(a)). As a quantitative measure of the interaction between the NA and the incident field, we employ the scattering cross section of the NA denoted by $\sigma_s$, defined as the total scattered power normalized to the incident field irradiance [56]. First, we employ a full wave simulation analysis based on the finite element method (FEM) implemented in CST Microwave Studio [57] to obtain the NA's $\sigma_s$, shown in Figure 1 (b), when it is irradiated by a plane wave with electric field $\mathbf{E}_{\text{inc}} = E_0 e^{ik_0 z}\hat{\mathbf{y}}$ versus wavelength within the interval of 600 nm to 800 nm, for various radii $a$. Here, $E_0$ is the complex amplitude of the electric field, and $\hat{\mathbf{x}}$ and $\hat{\mathbf{y}}$ are unit vectors of the Cartesian coordinate system. In this example in the following ones we assume that the high-density NA is made of Silicon, whose refractive index is extracted from Ref. [58].

To elaborate the essence of the interaction, we model the NA by induced electric $\mathbf{p}$ and magnetic $\mathbf{m}$ dipole moments and neglect the contributions from higher multipoles in the scattering. Such an approximate modeling is valid as long as the NA is optically small and has been widely used in metasurface design and analysis [59–62]. The dipole moments are related to the incident field $\mathbf{E}_{\text{inc}}^o$ and $\mathbf{H}_{\text{inc}}^o$ at the position of the NA through

$$\mathbf{p} = \alpha_{\text{ee}} \mathbf{E}_{\text{inc}}^o,$$

$$\mathbf{m} = \alpha_{\text{mm}} \mathbf{H}_{\text{inc}}^o. \qquad (5)$$

Here, $\alpha_{\text{ee}}$ and $\alpha_{\text{mm}}$ are electric and magnetic polarizabilities that are here evaluated using Mie theory. (For the sake of clarity, the magnetic dipole is defined here as $\mathbf{m} = \frac{1}{2}\int \mathbf{r} \times \mathbf{J}(\mathbf{r}) dv$ where $\mathbf{J}(\mathbf{r})$ represents the displacement current density.) In Figure 1 (b), we plot also the scattering cross section of a spherical NA with radius $a = 85$ nm calculated using only the induced electric and magnetic dipoles (continuous orange curve) showing that this representation leads to a good approximation.

In what follows, without loss of generality we focus on a NA with radius $a = 85$ nm made of Silicon with the peak of scattering cross section $\sigma_s$ occurring at wavelength $\lambda = 675$ nm which is almost in the middle of the considered wavelength range.

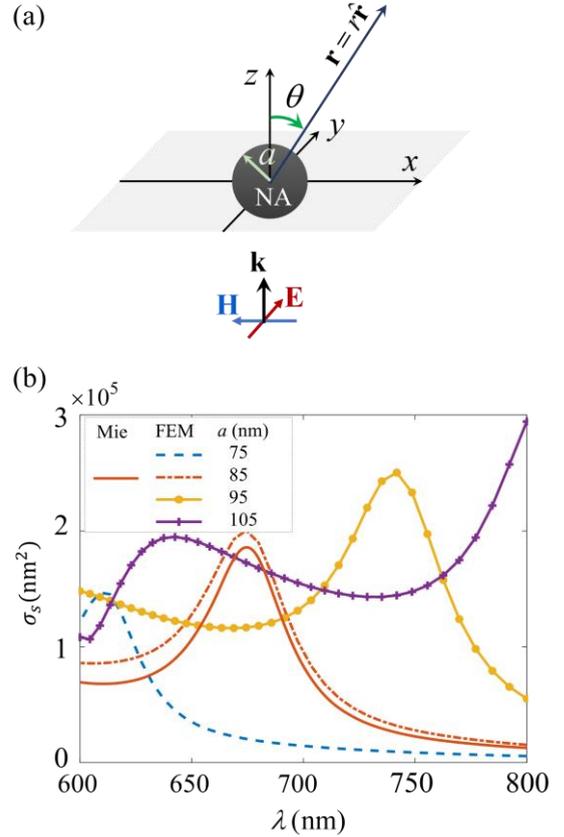

Figure 1. (a) High-density dielectric nanoantenna (NA) with radius $a$ located at the origin of a coordinate system and illuminated by a plane wave. (b) Scattering cross section $\sigma_s$ of the NA versus operating wavelength $\lambda$ for four distinct sphere's radii. In this example and in all the following ones the NA is made of Silicon. As it will be clear in the next figure, the peaks represent magnetic resonances.

The magnitudes of electric and magnetic dipole polarizabilities of a NA with radius $a = 85$ nm, and their phase difference $\Delta\varphi$ are calculated using the expressions from Mie theory and are plotted in Figure 2. This figure shows that at $\lambda = 730$ nm the electric and magnetic polarizabilities are approximately balanced, i.e., when $|\alpha_{\text{ee}}|/\epsilon_0 = |\alpha_{\text{mm}}|$ and their phase difference is very close to zero, which means that the Kerker condition [63–67] is almost satisfied at that wavelength and hence the NA directs the scattered field in the forward direction (a source with zero back radiation is often referred to as Huygens source [68–71]). The peak of $\alpha_{\text{mm}}$ at wavelength $\lambda = 675$ nm is often referred to as "magnetic resonance" of the NA.

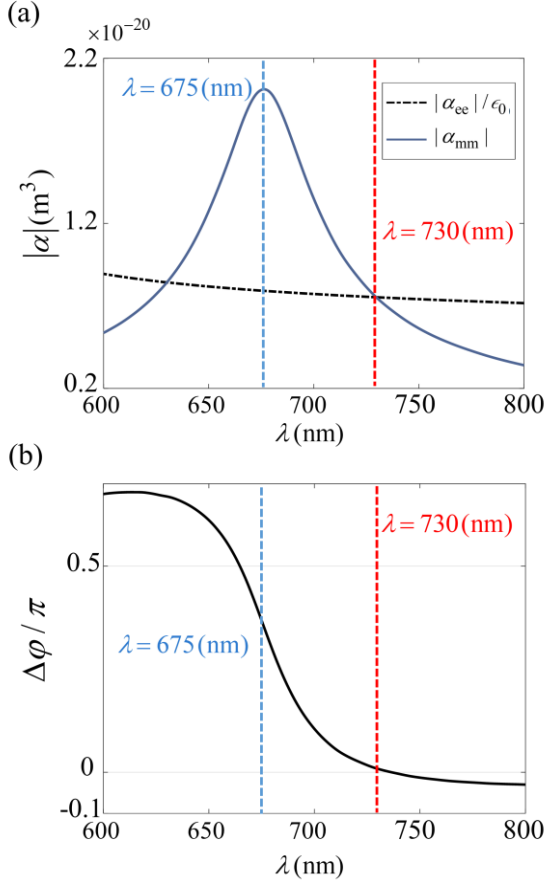

Figure 2. (a) Magnitude and (b) normalized phase difference between of the electric $\alpha_{ee}$ and magnetic $\alpha_{mm}$ dipole polarizabilities of a single NA with radius $a$=85 nm made of Silicon. At $\lambda = 730$ nm, the Kerker condition [63] is approximately satisfied, i.e., the balanced polarizabilities very closely satisfy $|\alpha_{ee}|/\epsilon_0 = |\alpha_{mm}|$ and $\Delta\varphi = 0$.

This result indicates that an array composed of NAs with radius $a = 85$ nm (to be investigated later in this paper) would mainly transmit all the incident power at $\lambda = 730$ nm which is advantageous when transmitted power is employed to detect chirality in accordance to standard circular dichroism experiments. However, coupling among the NAs (and a possible phase difference between incident and scattered fields) would cause a wavelength shift for maximum transmittance to occur.

Since the upper bound of helicity density is linearly linked to the energy density as was shown in Ref.[16], to have optimally chiral nearfield with maximized helicity density, it is required to simultaneously enhance electric and magnetic energy densities. However, according to Eq. (5), the induced electric **p** and magnetic **m** dipole moments are, respectively, oriented along the $y$ and $x$ axes when the NA is illuminated by $y$-polarized (electric field) plane wave. Therefore, electric energy density is enhanced more effectively along the $y$ axis while the maximum of magnetic energy density occurs along the $x$ axis. In the following subsection, we analyze helicity density of the total nearfield of a NA illuminated by a plane wave with circular polarization.

### A. Optimally chiral nearfield of a NA illuminated by a plane wave with circular polarization

We illuminate a single NA made of Si with radius $a$=85 nm by a left-handed circularly polarized plane wave with electric field $\mathbf{E}_{inc}$ represented as

$$\mathbf{E}_{inc} = E_0(\hat{\mathbf{x}} + i\hat{\mathbf{y}})e^{ik_0z}/\sqrt{2}, \quad (6)$$

propagating along the $+z$ axis and magnetic field $\mathbf{H}_{inc} = (E_0/\eta_0)(\hat{\mathbf{y}} - i\hat{\mathbf{x}})e^{ik_0z}/\sqrt{2}$ is obtained from Maxwell equation $\mathbf{H} = (i\omega\mu_0)^{-1}\nabla \times \mathbf{E}$. The wavenumber in vacuum is denoted by $k_0 = \omega\sqrt{\epsilon_0\mu_0}$. Here and in the following a hat tags unit vectors. It has been shown that helicity density $h$ of the total field around a NA is the sum of the helicity density of the incident field $h_{inc}$, the helicity density of the scattered field $h_{sca}$, and an interference helicity density $h_{int}$[36,55], i.e.,

$$h = h_{inc} + h_{int} + h_{sca}. \quad (7)$$

The sign of helicity density of the circularly polarized incident field is positive based on the definition in Eq. (2). Scattering helicity density $h_{sca}$ at location $\mathbf{r} = r\,\hat{\mathbf{r}}$, generated by a NA in its surrounding, when the NA's scattering is represented by only dipole moments **p** and **m** is approximated by the expression [36,55]

$$h_{sca} \approx \frac{\eta_0}{32\pi^2\omega r^6}\Im\{3(\hat{\mathbf{r}}\cdot\mathbf{p})(\hat{\mathbf{r}}\cdot\mathbf{m}^*) + \mathbf{p}\cdot\mathbf{m}^*\}. \quad (8)$$

after keeping only terms with dependence $r^{-3}$ and ignoring those which depend on $r^{-1}$ and $r^{-2}$. Introducing Eqs. (5) and (6) into Eq. (8) simplifies the expression of scattering helicity density $h_{sca}$ to

$$h_{sca} \approx \frac{\epsilon_0|E_0|^2}{32\pi^2\omega r^6}\left(\frac{3}{2}\sin^2\theta + 1\right)\Re\{\alpha_{ee}\alpha_{mm}^*\}/\epsilon_0, \quad (9)$$

where $\theta$ is the polar angle in spherical coordinate system $(r, \theta, \varphi)$. Figure 3 shows the term $\Re\{\alpha_{ee}\alpha_{mm}^*\}/\epsilon_0$ in Eq. (9) which relates the scattering helicity density to the electric and magnetic polarizabilities of the NA. Besides the slow varying $1/\omega$ term in Eq. (9), it is $\Re\{\alpha_{ee}\alpha_{mm}^*\}/\epsilon_0$ that controls the maximum of the scattering helicity density. Note that helicity enhancement, i.e., normalized scattering helicity density $h_{sca}$ to that of the incident field $h_{inc} = \epsilon_0|E_0|^2/(2\omega)$, is independent from amplitude and angular frequency of the excitation and is proportional to $h_{sca}/h_{inc} \propto \Re\{\alpha_{ee}\alpha_{mm}^*\}/\epsilon_0$. Therefore, since this term, as depicted in Figure 3, does not drop dramatically, the discussed NA provides enhanced helicity density over a wide wavelength range between 700 nm and 800 nm.

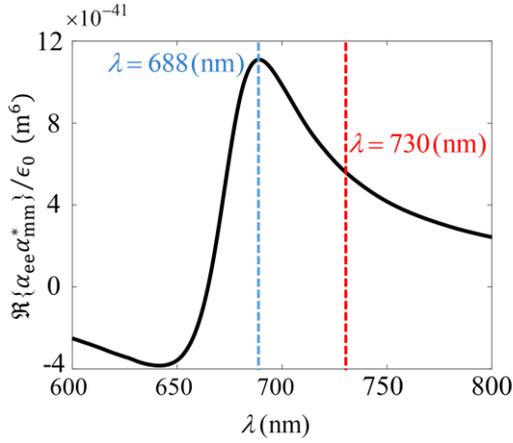

Figure 3. Term $\Re\{\alpha_{ee}\alpha_{mm}^*\}/\epsilon_0$, which is proportional to $h_{sca}/h_{inc}$, for a single NA with radius $a$=85nm versus wavelength. This term takes its maximum around $\lambda = 688$ nm, while it has a lower value at $\lambda = 730$ nm where polarizabilities are balanced.

Figure 3 shows that the scattering helicity density generated by a NA illuminated by a plane wave with CP reaches its maximum around $\lambda = 688$ nm while its value is smaller at $\lambda = 730$ nm where the two polarizabilities are balanced (when $\alpha_{ee}/\epsilon_0 = \alpha_{mm}$)). In Ref. [36,55] it is proved that balanced polarizabilities (when $\alpha_{ee}/\epsilon_0 = \alpha_{mm}$) are required to generate optimally chiral field around a NA, when it is illuminated by an optimally chiral incident field satisfying condition (3) such as a plane wave with circular polarization. This means that the maximum of helicity density does not corresponds to the "optimal chiral condition". However, for applications which require optimally chiral fields, such as chirality characterization of nanoparticles, this condition is satisfied at expense of losing a portion of helicity enhancement. Note that as discussed earlier in this paper, balanced polarizabilities also correspond to satisfaction of Kerker condition which is advantageous for applications such as CD experiments where measuring the transmitted power through an array of discussed NAs is the goal.

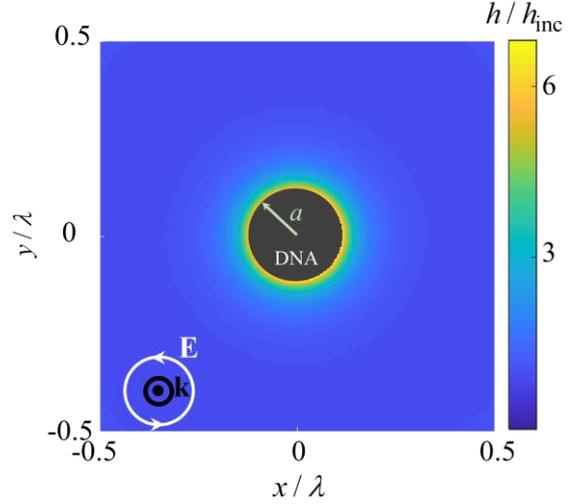

Figure 4. Normalized helicity density to that of the incident field around a single NA with radius $a$ =85 nm, irradiated by a plane wave with circular polarization at $\lambda = 730$ nm, in $xy$ plane, i.e., at $\theta = \pi/2$. Since illumination has circular polarization the distribution is symmetric in the azimuthal plane.

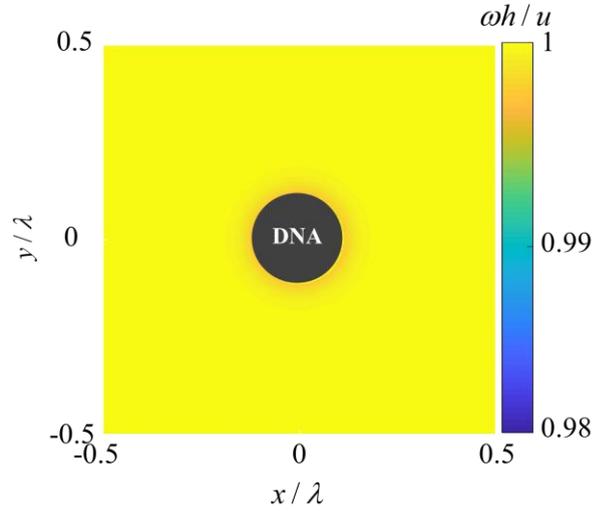

Figure 5. Plot of $\omega h/u$, i.e., of the product of the magnitude of helicity density and angular frequency normalized to the energy density, i.e., around a single NA with radius $a$=85 nm in the $xy$ plane at wavelength $\lambda$=730 nm. This quantity equals unity for optimally chiral fields and fields whose $\omega|h|/u$ values are close to unity are close to satisfying the required conditions to be optimally chiral.

Since the incident field has a left-handed circular polarization, the term $h_{inc}$ has positive value. Therefore, incident and scattering helicity densities interfere constructively contributing to a higher total helicity density at wavelength where $\Re\{\alpha_{ee}\alpha_{mm}^*\}/\epsilon_0$ takes positive values. Figure 4 illustrates the distribution of total helicity density (of the total field), obtained from full wave simulations, normalized to that of incident field in the transverse $xy$ plane, at $\lambda = 730$ nm.

According to Eq. (9), the helicity density of the scattered field $h_{\text{sca}}$ drops fast with a factor of $r^{-6}$ since it is mainly due to the nearfield. Moreover, as discussed in Ref. [36], the interference helicity density $h_{\text{int}}$, i.e., the second term in Eq. (9), drops by a factor of $r^{-3}$. Therefore, the total helicity density enhancement defined as $h/h_{\text{inc}}$ tends to unity as $1 + h_{\text{int}}/h_{\text{inc}} + h_{\text{sca}}/h_{\text{inc}}$ when moving away from the particle. Therefore, to provide enhanced optimally chiral nearfield with maximized helicity density generated by an individual NA (Figure 5), over a broader area, we propose an array of NAs where the *effective* polarizability of each NA in the array approximately satisfies the Kerker condition in the presence of mutual couplings among the NAs.

### IV. Optimally chiral nearfield of a planar Array of NAs

So far, we have shown that an individual achiral NA with balanced electric and magnetic polarizabilities $\alpha_{\text{ee}}/\epsilon_0 = \alpha_{\text{mm}}$ generates optimally chiral nearfield. In this section we extend this discussion to an *array* of NAs and determine the conditions required for an array to generate an *optimally chiral nearfield* over its surface. We investigate helicity density enhancement, maximization, and distribution over the surface of a planar array composed of NAs as in Figure 6, illuminated by a plane wave with circular polarization. Full wave simulations are carried out by using the frequency domain finite element method module in the software package CST Studio Suite by DS SIMULIA. Floquet boundary conditions are imposed on *x-y* boundaries of the array's unit cell and Floquet modes up to first order are considered above and below the array.

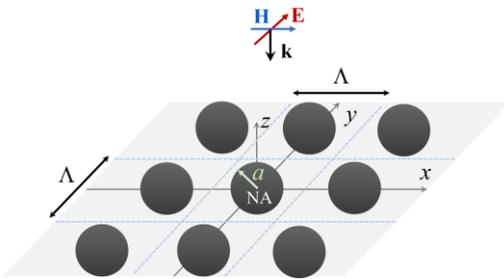

Figure 6. Array of NAs with radius *a* arranged in the *xy* plane with period Λ, illuminated by a plane wave, propagating along the negative *z* direction.

The properties of an array of NAs, located at $\mathbf{r}_{mn} = m\Lambda\hat{\mathbf{x}} + n\Lambda\hat{\mathbf{y}}$ with $m,n = 0, \pm 1, \pm 2, ...$ as that shown in Figure 6, are determined by the electric and magnetic polarizabilities of the NA (which are dictated by the NA's radius *a*) and the periodicity of the array Λ, which influences the interaction between array elements. Here we mainly examine arrays composed of NAs with radius $a$ =85 nm considered in Section III.

The total electric **E** and magnetic **H** fields in presence of the array of NAs are

$$\mathbf{E} = \mathbf{E}_{\text{inc}} + \sum_{m,n=-\infty}^{\infty} \mathbf{E}_{mn}$$

$$\mathbf{H} = \mathbf{H}_{\text{inc}} + \sum_{m,n=-\infty}^{\infty} \mathbf{H}_{mn}, \quad (10)$$

where $\mathbf{E}_{mn}$ and $\mathbf{H}_{mn}$ are the scattered electric and magnetic fields by the NA with its center located at $\mathbf{r}_{mn}$ and they are obtained either by full wave simulations or by using dyadic Green's functions. To have an *optimally chiral total nearfield* in proximity of the array plane, it is required to have: (i) an optimally chiral incident field i.e., $\mathbf{E}_{\text{inc}} = \pm i\eta_0 \mathbf{H}_{\text{inc}}$, and (ii) optimally chiral scattered field by each of array elements i.e., $\mathbf{E}_{mn} = \pm i\eta_0 \mathbf{H}_{mn}$. Indeed, under these conditions, the total electric **E** and magnetic **H** fields satisfy the relation

$$\mathbf{E} = \mathbf{E}_{\text{inc}} + \sum_{m,n=-\infty}^{\infty} \mathbf{E}_{mn}$$

$$= \pm i\eta_0 \mathbf{H}_{\text{inc}} \pm \sum_{m,n=-\infty}^{\infty} i\eta_0 \mathbf{H}_{mn}$$

$$= \pm i\eta_0 \mathbf{H}, \quad (11)$$

hence they are optimally chiral. For scattered electric and magnetic fields to satisfy $\mathbf{E}_{mn} = \pm i\eta_0 \mathbf{H}_{mn}$, it is required that electric $\mathbf{p}_{mn}$ and magnetic $\mathbf{m}_{mn}$ dipole moments induced in the $mn$th NA**,** satisfy condition

$$\mathbf{p}_{mn} = \pm \frac{i}{c_0} \mathbf{m}_{mn}, \quad \forall\, m,n. \quad (12)$$

Note that dipole moments in each NA are induced by the sum of the incident field and the scattered fields generated by all other array elements, it is possible to reduce condition (12) to

$$\alpha_{\text{ee}}^{\text{eff}}/\epsilon_0 = \alpha_{\text{mm}}^{\text{eff}} \quad (13)$$

when *effective* electric $\alpha_{\text{ee}}^{\text{eff}}$ and magnetic $\alpha_{\text{mm}}^{\text{eff}}$ polarizabilities are defined as

$$\mathbf{p}_{mn} = \alpha_{\text{ee}}^{\text{eff}} \mathbf{E}_{\text{inc}}^{o}$$

$$\mathbf{m}_{mn} = \alpha_{\text{mm}}^{\text{eff}} \mathbf{H}_{\text{inc}}^{o}. \quad (14)$$

The two effective polarizabilities of each NA arranged in an array are here calculated by employing Green's functions[72,73] (see Appendix). Note that due to the symmetry of the structure and to normally incident excitation, the in-plane effective polarizabilities of the square array of Si spheres are isotropic. Since here we deal only with normal incidence we then consider the two polarizabilities as scalars. Because of what demonstrated in (11) and referring to condition (13) as the effective Kerker condition for an array of NAs we now investigate if this condition is satisfied for an array of Si nanospheres.

We first check the occurrence of the effective Kerker condition by checking the vanishing of reflectivity when the array is illuminated by a plane wave with linear polarization and normal incidence. Once this is established, we will observe how the effective Kerker condition for the array illuminated by circular polarization will lead to almost optimal helicity density.

The power reflectivity $r$ and transmittivity $t$ coefficients of the array of NAs with radius $a = 85$ nm when the array is irradiated by a plane wave with linear polarization propagating along the negative $z$ direction, calculated by using full-wave simulations, for three distinct periodicity parameters $\Lambda$ is depicted in Figure 7. This figure shows that the array has almost total transmittivity at the wavelength where the effective Kerker condition is satisfied [63]. This does not occur at the wavelength at which the scattering cross section $\sigma_s$ of a single NA takes its maximum (see Figure 1), though it is not far from that.

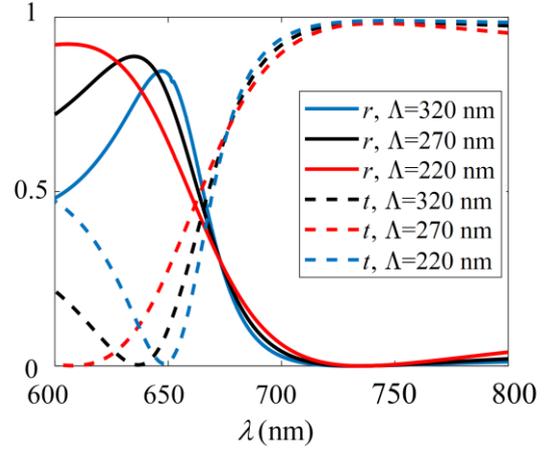

Figure 7. Power reflectivity $r$ and transmittivity $t$ of an array of high-density nanospheres with radius $a$=85 nm for three distinct periods $\Lambda$. Reflectivity $r$ is not very sensitive to change in periodicity $\Lambda$ and takes its minimum value very close to $\lambda = 730$ nm where Kerker condition [63] is satisfied for a single NA, and shifted only by a few nanometers.

As it is clear from this figure, the null in the reflectivity $r$ does not dramatically change with period length $\Lambda$ and it approximately (note a small shift in the wavelength of the null) occurs around a wavelength $\lambda = 730$ nm, where the Kerker condition [63] is satisfied for individual NA inclusions with radius $a$=85 nm. Indeed, our calculations show that effective polarizabilities of Si NA with radius $a$=85 nm arranged in an array[74] are slightly perturbed with respect to those of individual Si NA shown in Figure 2 and therefore changing periodicity slightly affects the wavelength at which the null in reflectivity occurs. Moreover, a zero transmission occurs at wavelengths in the range  600nm $< \lambda <$ 650 nm for the cases considered which shows that the second Kerker condition[66] is almost satisfied.

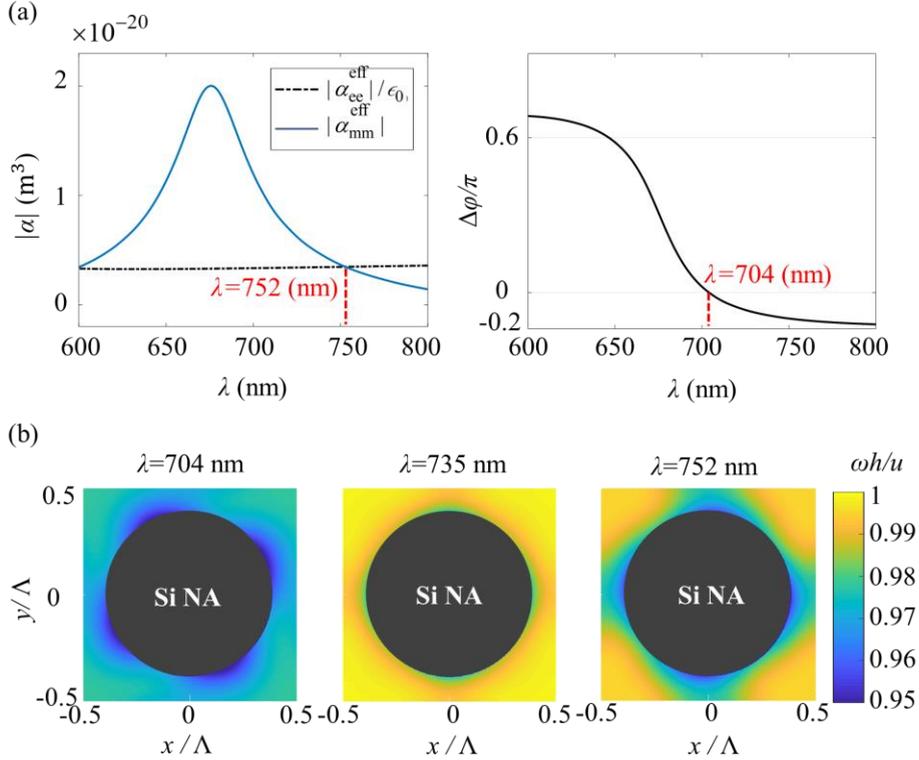

Figure 8. (a) Effective polarizabilities of an array of spherical Si NAs with radius 85 nm and periodicity $\Lambda = 220$ nm deviate form optimally chiral condition $\alpha_{ee}^{eff}/\epsilon_0 = \alpha_{mm}^{eff}$. (b) The quantity $h\omega/u$ in nearfield of the array illuminated by a left-handed circularly polarized plane wave at three different wavelengths: (i) at $\lambda = 704$ nm where $\Delta\varphi$ goes to zero; (ii), and at $\lambda = 735$ nm, where the minimum reflection from the array with periodicity $\Lambda = 220$ nm occurs; (iii) at $\lambda = 752$ nm where the relation $|\alpha_{ee}^{eff}|/\epsilon_0 = |\alpha_{mm}^{eff}|$ is satisfied. The relation $|h|\omega/u=1$ is most nearly satisfied at $\lambda = 735$ nm.

A. Helicity density of an array of NAs illuminated by a plane wave with circular polarization

As an illustrative example, the effective electric and magnetic polarizabilities of spherical Si NAs in an array with periodicity $\Lambda = 220$ nm are depicted in Figure 8(a). This figure shows the effective polarizabilities of Si NA deviates from the optimal chirality relation $\alpha_{ee}^{eff}/\epsilon_0 = \alpha_{mm}^{eff}$ since the phase difference $\Delta\varphi$ between effective polarizabilities does not vanish completely at the wavelength of 752 nm where the magnitudes of effective polarizabilities satisfy $|\alpha_{ee}^{eff}|/\epsilon_0 = |\alpha_{mm}^{eff}|$.

To investigate the closeness to the optimal chirality condition (3) in nearfield of this array, we illuminate the array with a left-handed circular polarized plane wave, and we plot in Figure 8 (b) the quantity $\omega h/u$ at the surface $z = 0$, at three different wavelengths: (i) at $\lambda = 704$ nm where $\Delta\varphi$ goes to zero; (ii) at $\lambda = 735$ nm, where the minimum reflection from the array with periodicity $\Lambda = 220$ nm occurs as shown in Figure 7; and (iii) at $\lambda = 752$ nm where relation $|\alpha_{ee}^{eff}|/\epsilon_0 = |\alpha_{mm}^{eff}|$ is satisfied.

Among the three cases the relation $|h|\omega/u=1$ is most nearly satisfied at $\lambda = 735$ nm. Therefore, this figure shows that the nearfield of the array is closest to satisfying optimally chiral condition where the minimum reflection occurs, which is between the two wavelengths where the magnitudes and phase difference of the effective polarizabilities satisfy the Kerker condition (13).

Note that lattice resonances, i.e., modes of the array, occur when the denominator of the effective polarizabilities (see Appendix) vanishes. This condition will be analyzed in a future investigation; however, we expect strong field and helicity enhancement close to these lattice resonances.

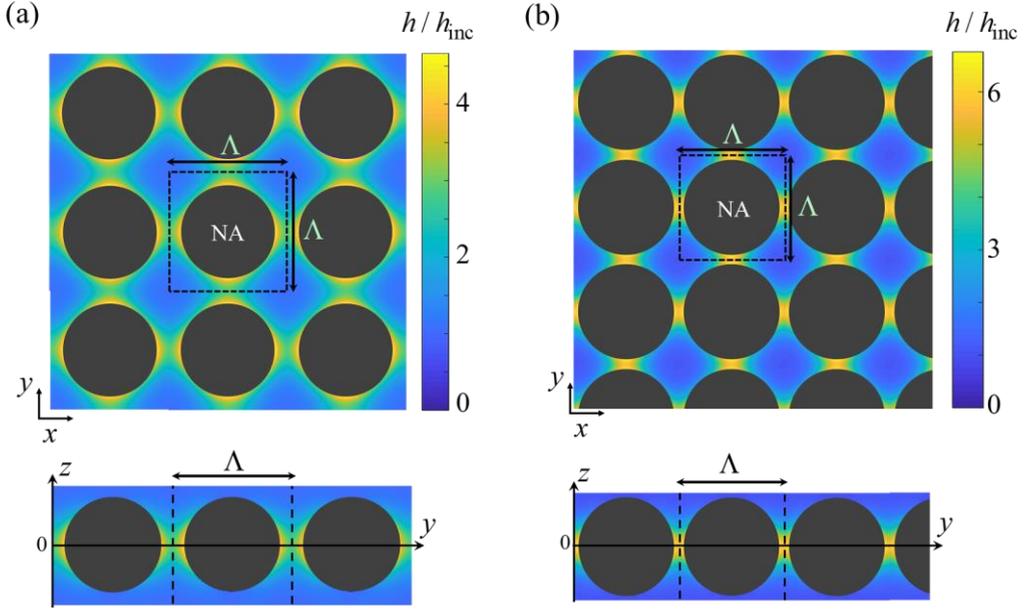

Figure 9. Helicity enhancement, i.e., helicity density of an array of high-density NAs, with radius $a = 85$nm made of Si, normalized to that of the incident left-handed circularly polarized field at wavelength $\lambda=735$ nm., evaluated at $z = 0$, and on the $yz$ plane, respectively. Two arrays are considered with period of (a) $\Lambda = 220$ nm, and (b) $\Lambda = 190$ nm. Note the different scale, and the higher helicity for the case with period $\Lambda = 190$ nm.

Next, we further analyze helicity density of the field in between the arrayed NAs at the specific wavelength corresponding to almost full transmission. As discussed earlier, when the Kerker condition (for a single NA) is satisfied, the reflectivity $r$ and transmittivity $t$ of the array are chiefly affected by the radius of the NAs, whereas the periodicity parameter $\Lambda$ does not significantly shift the maximum transmittivity's wavelength. However, when the periodicity parameter $\Lambda$ is reduced, the distance between NAs decreases. Therefore, energy density is localized in hot spots formed in spaces between NAs and the upper bound of helicity density which is linearly linked to energy density also increases.

In Figure 9, we show the total helicity density at the surface $z = 0$, normalized to that of the incident circularly polarized field, for two distinct array periods of $\Lambda = 220$ nm and $\Lambda = 190$ nm. This figure shows that the array with smaller periodicity parameter $\Lambda$ offers higher helicity enhancement in the hot spots formed between NAs. In particular, one can observe that for arrays with periodicity $\Lambda = 220$ nm and $\Lambda = 190$ nm such enhancement drops, respectively, to 80 and 90 percent of their values on the surface of the NA at a distance approximately equal to $0.015 \lambda$ away from the surface of the NA. Note that helicity enhancement around a *single* NA that was depicted in Figure 4 drops approximately to 75 and 50 percent of its value on the surface of the NA, respectively, at a distance equal to $0.015 \lambda$ and $0.03 \lambda$ away from the surface of the NA. To further clarify this point, in Figure 10, helicity enhancement is plotted varying position along the $x$ axis and along the diagonal line of a unit cell, for the array with periodicity $\Lambda = 190$ nm. This shows that an array of Si NA provides a helicity enhancement almost similar to that around a single NA, but with better uniform distribution. Indeed, the array arrangement of NAs spatially extends the enhanced helicity density over a large surface which is advantageous in some applications.

Note that the helicity densities illustrated in Figure 9 possess the same sign over the whole $z = 0$ surface and also off the $z = 0$ surface. Moreover, it is particularly important to have fields close to satisfying the optimally chiral condition $|h|\omega/u=1$ which maximizes helicity density at a given energy density.

As discussed in Section II, the field around a *single* NA with radius $a=85$ nm at wavelength $\lambda=730$ nm, are very close to be *optimally chiral* when helicity and energy densities satisfy Eq. (4). In Figure 11 we show the product of helicity density and angular frequency normalized to energy density, i.e., we show $\omega h/u$ (which equals unity for optimally chiral fields) at the surface $z=0$ of a unit cell of the array, in the $xy$ plane. Figure 11 shows that field over the surface of the array is very close to be optimally chiral, however, when the periodicity of the array decreases, the field deviates from the required condition (3) by approximately 5 per cent, so it is less optimally chiral.

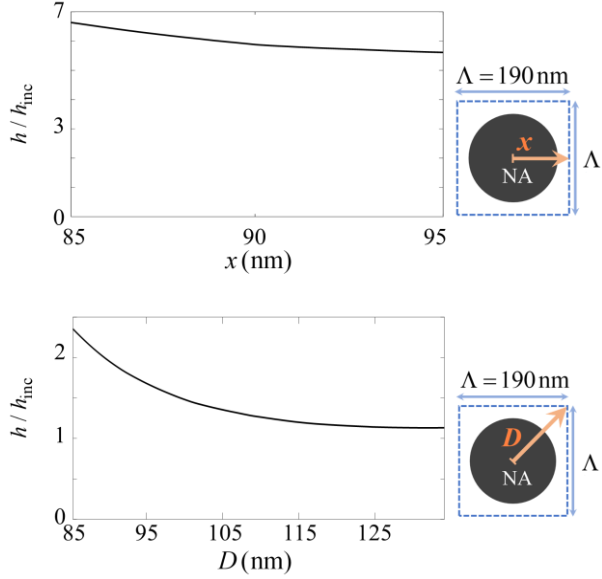

Figure 10. Helicity enhancement along the *x* axis and the diagonal line in a unit cell of an array of Si spheres with radius 85 nm and periodicity parameter $\Lambda = 190$, at the array plane $z = 0$.

### B. Effect of a substrate on helicity density generated by an array of Si NA

We examine here the impact of adding a glass substrate on the functionality of an array of high-density NAs. As an example, we consider an array of Si NAs with radius $a = 85$ nm and periodicity $\Lambda = 270$ nm located on top of a glass substrate with refractive index $n = 1.5$ and thickness $d$ depicted in Figure 12 (a). The substrate interface is at $z = -a$. We calculate the power reflectivity and transmittivity of this array for two cases: (a) when the substrate has an infinite thickness $d$ (it occupies the region $z < -a$) and (b) when the thickness is $d=980$ nm.

Figure 12 (b) represents the reflectivity $r$ and transmittivity $t$ of the array illuminated by a plane wave with circular polarization propagating along the negative $z$ direction. Note that the substrate with thickness $d=980$ nm fully transmits a normally incident plane wave at $\lambda = 735$ nm where the exact null of reflectivity $r$ of such an array of Si NAs occurs. This figure shows that when the substrate has a thickness $d=980$ nm, the wavelength at which the null of reflectivity $r$ occurs has changed very slightly (about 1 nm) with respect to the case when the substrate is not present at all (compare with Figure 7). However, for a substrate with an infinite thickness the reflectivity $r$ never approaches zero and takes a value approximately equal to 0.04 at $\lambda = 735$ nm.

Moreover, to investigate how the optimal chirality of the nearfield of the array is affected by the presence of the substrate, in Figure 13 we plot $\omega h/u$ over the $xy$ plane (at $z = 0$) in a unit cell of the array at wavelength $\lambda = 735$ nm. This figure illustrates that the nearfield over the surface of the array is very close to satisfying optimal chirality condition $\omega \frac{|h|}{u} = 1$ in the presence of the substrate with thickness $d=980$ nm. However, for a substrate with an infinite thickness this quantity decreases by 10 percent at wavelength $\lambda = 735$ nm which emphasizes the importance of optimizing the thickness of the substrate.

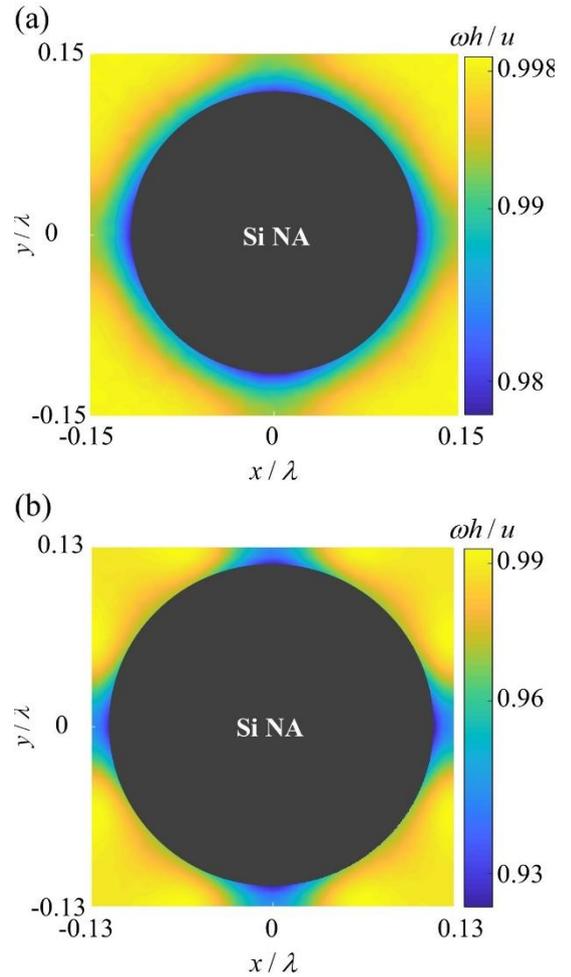

Figure 11. Plot of $\omega h/u$, i.e., the product of helicity density and angular frequency normalized to the energy density, generated by an array of high-density NAs with radius $a=85$ nm illuminated by a left-handed circularly polarized plane wave. We show such quantity in the $xy$ plane (at $z = 0$) over a unit cell of the array with period (a) $\Lambda = 220$ nm, and (b) $\Lambda = 190$ nm. Note that $\omega |h|/u = 1$ for optimally chiral fields, and this plot shows that the field in the NA array is close to that limit.

Therefore, the chiral operation of the array is weakly affected by the presence of the glass substrate when its thickness is optimized to transmit the total incident power. The small effect of the substrate for the optimized substrate thickness is because the NAs are mainly in air (above the substrate) and the contrast between the glass substrate and the air is not very high and hence the scattered nearfield by the NAs remains quite similar to the case without substrate.

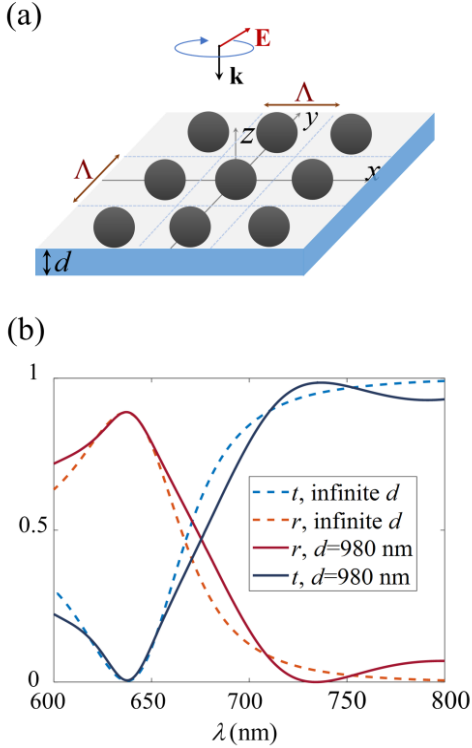

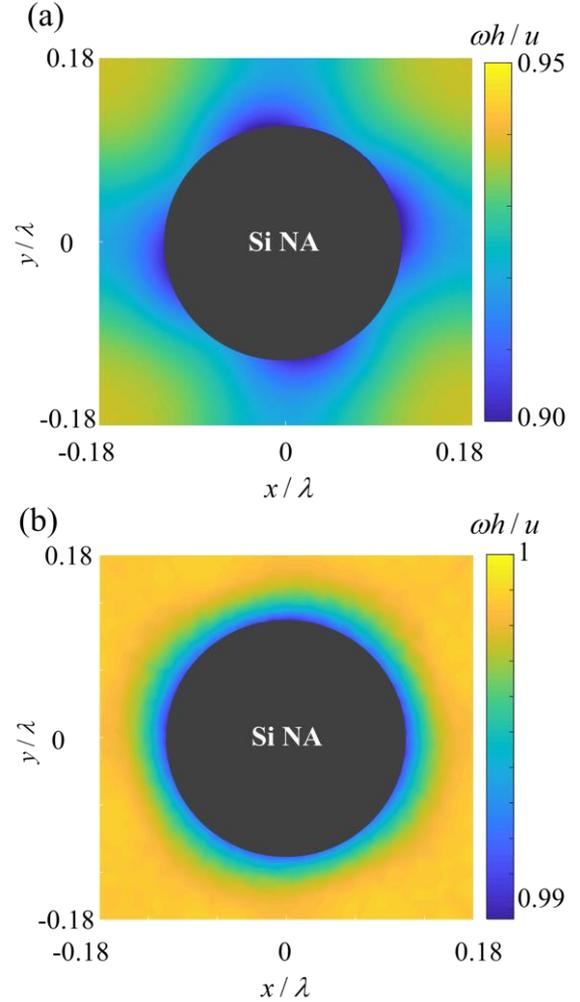

Figure 12. (a) An array of high-density NAs with radius $a$ =85 nm with periodicity $\Lambda$=270 nm located on a glass substrate with refractive index $n$ =1.5 illuminated by left-handed circularly polarized plane wave propagating along the negative $z$ direction, and (b) reflectivity $r$ and transmittivity $t$ of this array when the substrate has a finite thickness $d = 980$ nm and also for the semi-infinite case when it occupies the region $z < -a$.

Figure 13. Plot of $\omega h/u$, i.e., the product of helicity density and angular frequency normalized to the energy density, generated by an array of high-density NAs with radius $a$=85 nm and periodicity $\Lambda$=270 nm located on a glass substrate with refractive index $n$ =1.5 illuminated by a left-handed circularly wave at wavelength $\lambda = 735$ nm when (a) the substrate has an infinite thickness and it occupies the region $z < -a$ and (b) when it has a thickness $d = 980$ nm.

## V. CONCLUSION

We have shown that an array composed of spherical high-density dielectric nanoantennas illuminated by a plane wave with circular polarization generates nearfield very close to satisfying optimal chirality condition with enhanced helicity density. First, we have investigated helicity density in the nearfield of a single NA with balanced electric and magnetic polarizabilities, irradiated by a plane wave with circular polarization and discussed how the NA's scattered field is optimally chiral[16]. Helicity density is enhanced with a factor of 6.3 in vicinity of a single NA, symmetrically in the plane transverse to the propagation direction of the incident field. However, helicity density of the scattered fields drops quickly moving away from the surface of a single NA. Therefore, in the second part of the paper we have shown how helicity is enhanced over a surface of an array of NAs. We also show how the optimal chiral condition is related to the effective Kerker condition of the effective polarizabilities of the array particles and this is affected by the array period. Moreover, we investigated how the chiral operation of the array is affected in the presence of a glass substrate. The array nearfield slightly deviates from the optimal chiral condition when the inter-particle distances decreases but it generates a helicity enhancement over the whole array surface. This may present significant advantages in the detection of chirality of substances at surface level instead of in the bulk.

## ACKNOWLEDGMENTS

The authors are thankful for the support from the W. M. Keck Foundation, USA. The authors acknowledge fruitful discussions with Dr. Mohammad Albooyeh, University of California Irvine. Finally, the authors are thankful to DS SIMULIA for providing CST Studio Suite that was instrumental in this study.

## APPENDIX: EFFECTIVE POLARIZABILITIES OF NANOPARTICLES IN AN ARRAY

Electric $\mathbf{p}_{mn}$ and magnetic $\mathbf{m}_{mn}$ array dipole moments induced in the arrayed NAs with electric $\alpha_{ee}$ and magnetic $\alpha_{mm}$ polarizabilities located at $\mathbf{r}_{mn}$, are obtained as

$$\mathbf{p}_{mn} = \alpha_{ee} \mathbf{E}_{loc}(\mathbf{r}_{mn})$$

$$\mathbf{m}_{mn} = \alpha_{mm} \mathbf{H}_{loc}(\mathbf{r}_{mn}), \quad (A1)$$

where $\mathbf{E}_{loc}(\mathbf{r}_{mn})$ and $\mathbf{H}_{loc}(\mathbf{r}_{mn})$ are local electric and magnetic fields at the location of the $mn$th NA. Since the array is illuminated by a normally incident field, the electric and magnetic dipole moments in the array elements are all equal. Let us consider the reference NA at the origin of the coordinate with the local electric and magnetic fields

$$\mathbf{E}_{loc}(\mathbf{r}_{00}) = \mathbf{E}^o_{inc} + \breve{\underline{\mathbf{G}}}_{ee}(\mathbf{r}_{00}, \mathbf{r}_{00}) \cdot \mathbf{p}_{00},$$

$$\mathbf{H}_{loc}(\mathbf{r}_{00}) = \mathbf{H}^o_{inc} + \breve{\underline{\mathbf{G}}}_{mm}(\mathbf{r}_{00}, \mathbf{r}_{00}) \cdot \mathbf{m}_{00}, \quad (A2)$$

where $\breve{\underline{\mathbf{G}}}_{ee}(\mathbf{r}_{00}, \mathbf{r}_{00})$ and $\breve{\underline{\mathbf{G}}}_{mm}(\mathbf{r}_{00}, \mathbf{r}_{00})$ are the "regularized" dyadic Green's function that accounts for the field contributions produced by all the NAs except the one located at $\mathbf{r}_{00}$ as explained in Refs. [72,73]. Note that the electric and magnetic fields produced, respectively, by magnetic and electric dipole moments disappear due to the symmetry of the structure and its illumination. After substituting (A2) into (A1) and some algebraic manipulations, the induced dipole moments are obtained as

$$\mathbf{p}_{00} = \alpha_{ee} [\underline{\mathbf{I}} - \alpha_{ee} \breve{\underline{\mathbf{G}}}_{ee}(\mathbf{r}_{00}, \mathbf{r}_{00})]^{-1} \cdot \mathbf{E}^o_{inc},$$

$$\mathbf{m}_{00} = \alpha_{mm} [\underline{\mathbf{I}} - \alpha_{mm} \breve{\underline{\mathbf{G}}}_{mm}(\mathbf{r}_{00}, \mathbf{r}_{00})]^{-1} \cdot \mathbf{H}^o_{inc}, \quad (A3)$$

where $\underline{\mathbf{I}}$ is the identity dyadic. Therefore, the effective polarizabilities are defined as

$$\underline{\boldsymbol{\alpha}}^{eff}_{ee} = \alpha_{ee} (\underline{\mathbf{I}} - \alpha_{ee} \breve{\underline{\mathbf{G}}}_{ee}(\mathbf{r}_{00}, \mathbf{r}_{00}))^{-1},$$

$$\underline{\boldsymbol{\alpha}}^{eff}_{mm} = \alpha_{mm} (\underline{\mathbf{I}} - \alpha_{mm} \breve{\underline{\mathbf{G}}}_{mm}(\mathbf{r}_{00}, \mathbf{r}_{00}))^{-1}. \quad (A4)$$

Since the planar array has a square lattice, i.e., its periodicity along $x$ and $y$ directions are equal, only the diagonal components $\breve{G}_{ee,xx} = \breve{G}_{ee,yy}$, $\breve{G}_{mm,xx} = \breve{G}_{mm,yy}$, $\breve{G}_{ee,zz}$, and $\breve{G}_{mm,zz}$ of the Dyadic Green's functions shall be considered. Moreover, for a normally incident plane wave, only $\alpha^{eff}_{ee,xx}$, $\alpha^{eff}_{ee,yy}$, $\alpha^{eff}_{mm,xx}$, and $\alpha^{eff}_{mm,yy}$ are excited. Hence, the effective polarizabilities simplify to

$$\underline{\boldsymbol{\alpha}}^{eff}_{ee} = \alpha^{eff}_{ee} (\hat{\mathbf{x}}\hat{\mathbf{x}} + \hat{\mathbf{y}}\hat{\mathbf{y}})$$

$$\underline{\boldsymbol{\alpha}}^{eff}_{mm} = \alpha^{eff}_{mm} (\hat{\mathbf{x}}\hat{\mathbf{x}} + \hat{\mathbf{y}}\hat{\mathbf{y}}), \quad (A5)$$

where $\alpha^{eff}_{ee} = \alpha_{ee}/(1 - \alpha_{ee} \breve{G}_{ee,xx})$ and $\alpha^{eff}_{mm} = \alpha_{mm}/(1 - \alpha_{mm} \breve{G}_{mm,xx})$.